\documentclass[aps,prc,superscriptaddress,twoside,twocolumn,showpacs]{revtex4}
\usepackage{amsmath,amssymb}
\usepackage{graphicx}
\usepackage{graphics}
\usepackage{epsfig}

\begin{document}

\title{The production of $\boldsymbol{K^+K^-}$ pairs in proton-proton collisions at 2.83~GeV}
\author{Q.~J.~Ye}\email{qy4@phy.duke.edu}
\affiliation{Department of Physics and Triangle Universities Nuclear Laboratory, Duke University, Durham, NC 27708, USA}
\affiliation{Institut f\"ur Kernphysik and J\"ulich Centre for Hadron
Physics, Forschungszentrum J\"ulich, D-52425 J\"ulich, Germany}
\author{M.~Hartmann}\email{m.hartmann@fz-juelich.de}
\affiliation{Institut f\"ur Kernphysik and J\"ulich Centre for Hadron
Physics, Forschungszentrum J\"ulich, D-52425 J\"ulich, Germany}
\author{Y.~Maeda}
\affiliation{Research Center for Nuclear Physics, Osaka
University, Ibaraki, Osaka 567-0047, Japan}
\author{S.~Barsov}
\affiliation{High Energy Physics Department, Petersburg Nuclear
Physics Institute, RU-188350 Gatchina, Russia}
\author{M.~B\"uscher}
\affiliation{Institut f\"ur Kernphysik and J\"ulich Centre for Hadron
Physics, Forschungszentrum J\"ulich, D-52425 J\"ulich, Germany}
\author{D.~Chiladze}
\affiliation{Institut f\"ur Kernphysik and J\"ulich Centre for Hadron
Physics, Forschungszentrum J\"ulich, D-52425 J\"ulich, Germany}
\affiliation{High Energy Physics Institute, Tbilisi State University, GE-0186
Tbilisi, Georgia}
\author{S.~Dymov}
\affiliation{Physikalisches Institut, Universit{\"a}t
Erlangen-N\"urnberg, D-91058 Erlangen, Germany}
\affiliation{Laboratory of Nuclear Problems, Joint Institute for
Nuclear Research, RU-141980 Dubna, Russia}
\author{A.~Dzyuba}
\affiliation{High Energy Physics Department, Petersburg Nuclear
Physics Institute, RU-188350 Gatchina, Russia}
\author{H.~Gao}
\affiliation{Department of Physics and Triangle Universities Nuclear Laboratory, Duke University, Durham, NC 27708, USA}
\author{R.~Gebel}
\affiliation{Institut f\"ur Kernphysik and J\"ulich Centre for Hadron
Physics, Forschungszentrum J\"ulich, D-52425 J\"ulich, Germany}
\author{V.~Hejny}
\affiliation{Institut f\"ur Kernphysik and J\"ulich Centre for Hadron
Physics, Forschungszentrum J\"ulich, D-52425 J\"ulich, Germany}
\author{A.~Kacharava}
\affiliation{Institut f\"ur Kernphysik and J\"ulich Centre for Hadron
Physics, Forschungszentrum J\"ulich, D-52425 J\"ulich, Germany}
\author{I.~Keshelashvili}
\affiliation{Department of Physics, University of Basel, CH-4056 Basel,
Switzerland}
\author{Yu.~T.~Kiselev}
\affiliation{Institute for Theoretical and Experimental Physics,
RU-117218 Moscow, Russia}
\author{A.~Khoukaz}
\affiliation{Institut f\"ur Kernphysik, Universit\"at
M\"unster, D-48149 M\"unster, Germany}
\author{V.~P.~Koptev}\thanks{Deceased}
\affiliation{High Energy Physics Department, Petersburg Nuclear
Physics Institute, RU-188350 Gatchina, Russia}
\author{P.~Kulessa}
\affiliation{H.~Niewodniczanski Institute of Nuclear Physics PAN,
PL-31342 Cracow, Poland}
\author{A.~Kulikov}
\affiliation{Laboratory of Nuclear Problems, Joint Institute for
Nuclear Research, RU-141980 Dubna, Russia}
\author{B.~Lorentz}
\affiliation{Institut f\"ur Kernphysik and J\"ulich Centre for Hadron
Physics, Forschungszentrum J\"ulich, D-52425 J\"ulich, Germany}
\author{T.~Mersmann}
\affiliation{Institut f\"ur Kernphysik, Universit\"at
M\"unster, D-48149 M\"unster, Germany}
\author{S.~Merzliakov}
\affiliation{Institut f\"ur Kernphysik and J\"ulich Centre for Hadron
Physics, Forschungszentrum J\"ulich, D-52425 J\"ulich, Germany}
\affiliation{Laboratory of Nuclear Problems, Joint Institute for
Nuclear Research, RU-141980 Dubna, Russia}
\author{S.~Mikirtytchiants}
\affiliation{Institut f\"ur Kernphysik and J\"ulich Centre for Hadron
Physics, Forschungszentrum J\"ulich, D-52425 J\"ulich, Germany}
\affiliation{High Energy Physics Department, Petersburg Nuclear
Physics Institute, RU-188350 Gatchina, Russia}
\author{M.~Nekipelov}
\affiliation{Institut f\"ur Kernphysik and J\"ulich Centre for Hadron
Physics, Forschungszentrum J\"ulich, D-52425 J\"ulich, Germany}
\author{H.~Ohm}
\affiliation{Institut f\"ur Kernphysik and J\"ulich Centre for Hadron
Physics, Forschungszentrum J\"ulich, D-52425 J\"ulich, Germany}
\author{E.~Ya.~Paryev}
\affiliation{Institute for Nuclear Research, Russian Academy of
Sciences, RU-117312 Moscow, Russia}
\author{A.~Polyanskiy}
\affiliation{Institut f\"ur Kernphysik and J\"ulich Centre for Hadron
Physics, Forschungszentrum J\"ulich, D-52425 J\"ulich, Germany}
\affiliation{Institute for Theoretical and Experimental Physics,
RU-117218 Moscow, Russia}
\author{V.~Serdyuk}
\affiliation{Institut f\"ur Kernphysik and J\"ulich Centre for Hadron
Physics, Forschungszentrum J\"ulich, D-52425 J\"ulich, Germany}
\affiliation{Laboratory of Nuclear Problems, Joint Institute for
Nuclear Research, RU-141980 Dubna, Russia}
\author{H.~J.~Stein}
\affiliation{Institut f\"ur Kernphysik and J\"ulich Centre for Hadron
Physics, Forschungszentrum J\"ulich, D-52425 J\"ulich, Germany}
\author{H.~Str\"oher}
\affiliation{Institut f\"ur Kernphysik and J\"ulich Centre for Hadron
Physics, Forschungszentrum J\"ulich, D-52425 J\"ulich, Germany}
\author{S.~Trusov}
\affiliation{Institut f\"ur Kern- und Hadronenphysik,
Helmholtz-Zentrum Dresden-Rossendorf, D-01314 Dresden, Germany}
\affiliation{Skobeltsyn Institute of Nuclear Physics, Lomonosov Moscow
State University, RU-119991 Moscow, Russia}
\author{Yu.~Valdau}
\affiliation{Institut f\"ur Kernphysik and J\"ulich Centre for Hadron
Physics, Forschungszentrum J\"ulich, D-52425 J\"ulich, Germany}
\affiliation{Helmholtz-Institut f\"ur Strahlen- und Kernphysik, Universit\"at
Bonn, D-53115 Bonn, Germany}
\author{C.~Wilkin}
\affiliation{Physics and Astronomy Department, UCL, London WC1E 6BT,
United Kingdom}
\author{P.~W\"ustner}
\affiliation{Zentralinstitut f\"ur Elektronik,
Forschungszentrum J\"ulich, D-52425 J\"ulich, Germany}
\date{\today}

\begin{abstract}
Differential and total cross sections for the $pp\to ppK^+K^-$ reaction have
been measured at a proton beam energy of 2.83~GeV using the COSY-ANKE
magnetic spectrometer. Detailed model descriptions fitted to a variety of
one-dimensional distributions permit the separation of the $pp\to pp\phi$
cross section from that of non-$\phi$ production. The differential spectra
show that higher partial waves represent the majority of the $pp\to pp\phi$
total cross section at an excess energy of 76~MeV, whose energy dependence
would then seem to require some $s$-wave $\phi p$ enhancement near threshold.
The non-$\phi$ data can be described in terms of the combined effects of
two-body final state interactions using the same effective scattering
parameters determined from lower energy data.
\end{abstract}
%
\maketitle
%
%
\section{Introduction}
\label{introduction}%

The phenomenological description of strangeness production in nucleon-nucleon
collisions near threshold is complicated for a variety of reasons and these
add to its general interest. The large mass changes involved are necessarily
associated with short-range phenomena and therefore stress the importance of
heavy meson exchange. Whether such exchanges are mainly of strange or
non-strange nature is still an open question.

It is known that the scattering lengths in the $\Lambda p$ and $K^- p$
systems are both very large and these will distort any spectra. Furthermore,
important transitions, such as $\Sigma p \rightleftarrows \Lambda p$, mean
that several of the final channels are strongly coupled. It is also possible
that, due to such a coupling, the production of a heavy hyperon in $pp\to
K^+p\Lambda(1405)$ might influence kaon pair production in the $pp\to
ppK^+K^-$ reaction~\cite{Wil09}. The only hope of being able to disentangle
such effects is through having detailed experimental spectra in different
kinematic variables.

In addition to explicit strangeness production one has to consider also
hidden strangeness, such as that residing in the $\phi$ meson which, in the
quark model, is mainly composed of $s\bar{s}$ pairs. The $pp\to pp\phi$ cross
section might therefore be influenced by other strangeness production
channels that are important at this energy. We know from the
Okubo-Zweig-Iizuka (OZI) rule~\cite{OZI} that the $\phi$ production rate
should be much lower than that of the $\omega$ meson in the $pp\to pp\omega$
channel. Some of the widely observed violations of this rule, such as for
example in the $\phi/\omega$ ratio measured in $p\bar{p}$
annihilation~\cite{Ams98}, might be understood if there were significant
strangeness components in the nucleon. There are, however, alternative
explanations in terms of modified meson exchange models~\cite{Loc95,Mei97}.
The production of the $\phi$ meson in proton-nucleus collisions will clearly
depend on the more elementary $pp\to pp\phi$ cross section and so this will
also be an important ingredient in the understanding of proton-induced
nuclear transparency measurements~\cite{Pol11}.

The production of the $\phi$ meson in $pp$ collisions has been studied in
several theoretical papers within a simple one-meson-exchange
model~\cite{Sib96}, including both mesonic and nucleonic currents
components~\cite{Tit99,Nak99,Tsu03,Kap05}, and also contributions from
nucleon resonances~\cite{Fae03,Xie08,Sib06}. However, the rather limited
published data set is not sufficient to provide strong constraints on the
different models. It is the aim of the present paper to present detailed
measurements of the $pp\to ppK^+K^-$ reaction at a beam energy of 2.83~GeV,
where the cross sections for the production of the $\phi$ meson is cleanly
separated from the non-$\phi$.

We have previously published data on the $pp\to ppK^+K^-$
reaction~\cite{Har06,Mae08} at excess energies with respect to the $\phi$
threshold, $\varepsilon_{\phi}=\sqrt{s}-(2m_p+m_{\phi})c^2$, of 18.5, 34.5,
and 76~MeV, where $\sqrt{s}$ is the total center-of-mass (c.m.) energy. The
data on $\phi$ production at the lowest energy are consistent with the
particles in the final state being all in relative $S$-waves, with the only
feature evident in the measured spectra coming from the strong proton-proton
final state interaction (FSI). The lower statistics at the two higher
energies were sufficient to extract total cross sections but it was hard to
draw firm conclusions regarding the differential spectra which, on general
grounds, are expected to be much richer than at $\varepsilon_{\phi} =
18.5$~MeV.

In the pioneering work of the DISTO collaboration, strong evidence was
presented for the importance of higher partial waves in the $pp\to pp\phi$
reaction at $\varepsilon_{\phi}=83$~MeV~\cite{Bal01}, but no attempt was made
to make a consistent partial wave decomposition. The reason was in part due
to the necessity to study in detail the structure of the non-$\phi$ $K^+K^-$
background. We have since then shown~\cite{Mae08} that, due to the $K^-p$
final state interaction, the differential spectra for the non-$\phi$
contribution to the $pp\to ppK^+K^-$ reaction are strongly distorted. As a
consequence, one needs full descriptions of both $\phi$ and non-$\phi$
components in order to extract credible partial wave parameters. When these
are implemented in our current data set it is found that only a small amount
of the total $pp\to pp\phi$ cross section at $\varepsilon_{\phi}=76$~MeV
corresponds to pure $S$-wave final states and this explains the
non-observation of the $S$-wave $pp$ FSI enhancement in the results. However,
the energy dependence of the total $\phi$ production cross section then seems
to require some enhancement in the $\phi p$ system at low invariant masses.

The distortion of the $K^-p$ and $K^-pp$ invariant mass spectra observed in
the $pp\to ppK^+K^-$ reaction were very well parametrized by assuming
factorized pair-wise final state interactions in the $pp$, $K^-p$, and
$K^+K^-$ systems with constant effective scattering lengths~\cite{Mae08}.
Furthermore, their inclusion led to a good description of the energy
dependence of the total cross section, including the low energy COSY-11
data~\cite{Wol98,Que01,Win06}. The distortions were in fact first identified in
data taken below the $\phi$ threshold where selection of the non-$\phi$
contribution is automatic~\cite{Win06}. These distortions were well described
by the FSI parameters used at higher energies but, taken in isolation, the
error bars on the parameters extracted from these low energy data were very
large~\cite{Sil09}.

We here present much more precise differential data for the $pp\to ppK^+K^-$
reaction at a beam energy of $T_p= 2.83$~GeV ($\varepsilon_{\phi} = 76$~MeV)
obtained using the COSY-ANKE spectrometer. These will challenge the
theoretical models that can describe well the energy dependence of the $pp\to
pp\phi$ total cross sections. Currently few of the experimental spectra are
calculated in any of the phenomenological approaches.

The paper is organized as follows. We first describe the experimental setup
and data analysis in Sec.~\ref{Experiment}. The detailed phenomenological
parametrizations developed for $\phi$ and non-$\phi$ $K^+K^-$ production
needed to make the acceptance corrections are described here. The resulting
differential distributions for $\phi$ production are presented in
Sec~\ref{phiresults}, with the integrated cross sections for all the $pp\to
ppK^+K^-$ data being given in Sec.~\ref{total}. The distortions in the
non-$\phi$ differential cross sections arising from the various final state
interactions are discussed in Sec.~\ref{nonphi}, followed by our conclusions
in Sec.~\ref{conclusions}.

%
%
\section{Experiment and data analysis}
\label{Experiment}%

The experiment was performed at the Cooler Synchrotron (COSY) of the
Forschungszentrum J\"{u}lich ~\cite{Mai97} using the ANKE magnetic
spectrometer~\cite{Bar01,Har07} that is located at an internal target station
of the storage ring. ANKE contains three dipole magnets; \emph{D1} and
\emph{D3} divert the circulating beam onto the target and back into the COSY
ring, respectively, while \emph{D2} is the analyzing magnet. There are
detection systems placed to the right and left of the beam that register
positively and negatively charged ejectiles, respectively, while fast
positive particles are measured in the forward detector. Both the positive
and negative side detectors consist of sets of start and stop scintillation
counters for time-of-flight (TOF) measurements and two Multiwire Proportional
Chambers (MWPCs) utilized for particle momentum reconstruction. Two layers of
scintillation hodoscopes and three MWPCs are incorporated in the forward
detector which, in addition to studying a fast proton from the $pp\to
ppK^+K^-$ reaction, is also used to measure subsidiary reactions that are
needed to determine the luminosity~\cite{Bar01,Dym04,Har06}. A high density
hydrogen cluster-jet target~\cite{Kho99} was employed in the experiment.

Particle identification relied on momentum determination and time-of-flight
measurements~\cite{Har06,Mae08,Bus02}. The time differences were calibrated
by using, respectively, $\pi^+\pi^-$ and $\pi^+p$ pairs for the negative and
forward STOP counters with respect to the positive STOP counters, described
in detail in Ref.~\cite{Har07}. The kaon detection efficiency depends on the
particle's momentum and varied between 92\% and 98\%, whereas that for the
forward-going protons was about 96\%. The uncertainties in the efficiency
estimates were about 3\%.

The $pp\to ppK^{+}K^{-}$ reaction was identified through a triple-coincidence
involving the detection of a $K^+K^-$ pair and a forward-going proton, with
the additional requirement that the missing mass of the $K^+K^-p$ system be
consistent with that of the proton. In the off-line analysis, positive kaons
were selected by a dedicated $K^+$ detection system using the TOF information
between the START and the STOP counters~\cite{Bus02}. The $K^-$ and
forward-going protons were then selected simultaneously using the
time-of-flight differences, as described in detail in Ref.~\cite{Har07}.

\begin{figure}[h!]
\centering
\includegraphics[width=0.9\columnwidth,clip]{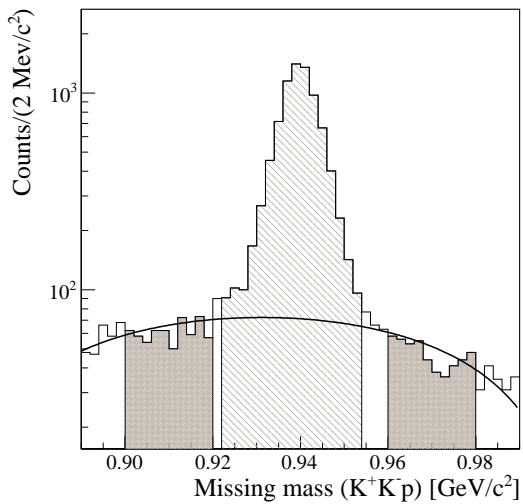}
\caption{The $K^+K^-p$ missing-mass distribution in the $pp\to pK^+K^-X$
reaction at $T_p=2.83$~GeV. The hatched histogram shows the cuts used for the
selection of the non-detected proton. The solid line, which is a second-order polynomial curve, 
was used in the analysis to estimate the background contribution under the proton peak.} \label{MM2.83}
\end{figure}

All time-of-flight selections, as well as the identification of the $K^+$,
were performed within $\pm3\sigma$ ranges. A similar cut was also made in the
missing-mass distribution of the detected $K^+K^-p$ shown in
Fig.~\ref{MM2.83}. The fraction of misidentified events inside the
$\pm3\sigma$ ($\sigma=4.7$~MeV/$c^2$) cut window around the proton mass was
estimated to be about 11.5\%, which was subtracted in the analysis using
weighted data from the side bands, as parametrized by the solid line. Any
ambiguity in this procedure is less than 3\% and is considered as one source
of systematic uncertainties in the analysis.

\begin{figure}[h!]
\centering
\includegraphics[width=0.9\columnwidth,clip]{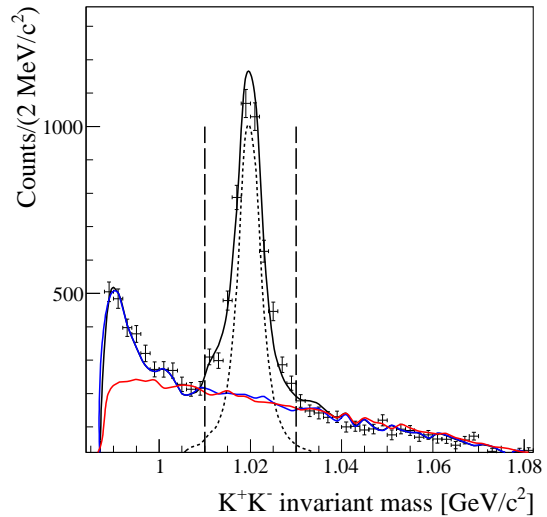}
\caption{(Color online) The raw $K^+K^-$ invariant mass distribution, IM$_{K^+K^-}$ (points),
is compared with the distribution of events obtained in a Monte Carlo
simulation (curve). The error bars indicate only the statistical
uncertainties. The blue curve shows the non-$\phi$ contributions within the
fitted parametrization, the red curve the four-body phase-space simulation of
$ppK^+K^-$, and the dotted histogram the $\phi$ contributions. The solid line
is the incoherent sum of the $\phi$ and non-$\phi$ contributions. The vertical
lines indicate the cuts used for the separation of the $\phi$-rich and non-$\phi$
regions. The fluctuations reflect the Monte Carlo sampling effects.} \label{IM}
\end{figure}

Having identified good $pp\to ppK^+K^-$ events, these were binned in terms of
the $K^+K^-$ invariant mass, IM$_{K^+K^-}$, and the corresponding results are
shown in Fig.~\ref{IM}. A clear $\phi$ peak is observed above a slowly
varying background. The experimental data were then divided into two samples,
a $\phi$-rich region where
$1.01~\textrm{GeV}/c^2<\textrm{IM}_{K^+K^-}<1.03$~GeV/$c^2$ and a non-$\phi$
(the rest) region. The model-independent acceptance estimate method used in
our earlier work~\cite{Har06,Mae08}, cannot be applied in the present
analysis since, at this higher excess energy, the number of zero elements in
the acceptance matrix is significant and this leads to large fluctuations.
Phenomenological parametrizations that describe well the experimental data in
both the $\phi$ and non-$\phi$ regions must therefore be relied upon in order
to perform the necessary acceptance corrections.

We start the analysis with the kaon pair production away from the $\phi$
region since it is crucial to master this contribution to understand the
background under the $\phi$ peak. The ansatz in our previous work on
non-$\phi$ production~\cite{Mae08} was taken as the basis of the simulation.
Here it was assumed that the overall enhancement factor was the product of
enhancements in the $pp$ and two $K^-p$ systems:
\begin{equation}
\label{assume}
F = F_{pp}(q_{pp}) \times F_{Kp}(q_{Kp_{1}}) \times F_{Kp}(q_{Kp_{2}})
\end{equation}
where $q_{pp}$, $q_{Kp_{1}}$ and $q_{Kp_{2}}$ are the magnitudes of the
relative momenta in the $pp$ and the two $K^-p$ systems, respectively. Note
that it is believed that the $K^+p$ interaction might be weakly repulsive
and, if so, its effects would be interpreted as extra attraction in the
$K^-p$ system.

Using an effective $K^-p$ scattering length of
$a_{K^-p}=(0+1.5i)$~fm~\cite{Mae08}, together with an additional weight of
$1+2.0\cos^2\theta$ on the polar angle of the $K^+K^-$ system in the overall
c.m.\ system, the invariant mass distributions can be described very well,
except for the very low $K^+K^-$ invariant masses,
$\textrm{IM}_{K^+K^-}<995$~MeV/$c^2$. In this region there are small residual
effects associated with $K\bar{K}$ final state interactions~\cite{Dzy08}, to
which we shall return later.

\begin{figure}[h!]
\centering
\includegraphics[width=1.05\columnwidth,clip]{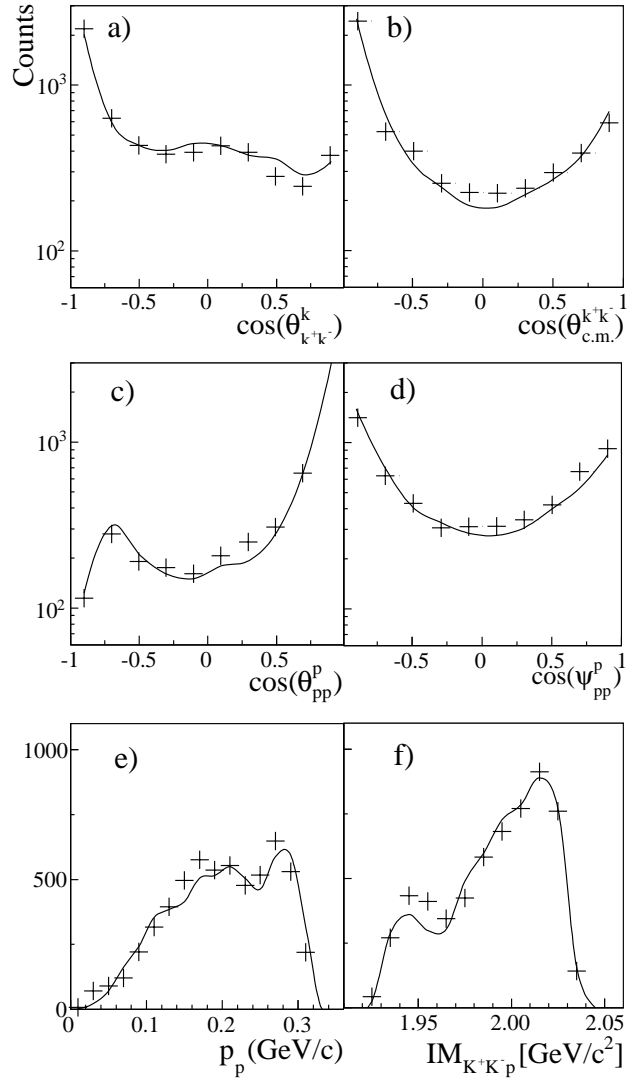}
\caption{Differential distributions of experimental (points) and simulated
(curves) yields for kaon pair production in the $pp \to ppK^+K^-$ reaction at
$T_p=2.83$~GeV for the non-$\phi$ regions
($\textrm{IM}_{K^+K^-}<1.01$~GeV/$c^2$ or
$\textrm{IM}_{K^+K^-}>1.03$~GeV/$c^2$). Vertical bars represent the
statistical uncertainties and horizontal ones the bin widths. The individual
panels are (a) the cosine of the polar angle of the $K^+$ in the $K^+K^-$
reference frame, (b) the polar angle of the kaon pairs in the overall c.m.\
frame, (c) the polar angle of the emitted proton in the $pp$ reference frame
relative to the beam direction, (d) the polar angle of the proton in the $pp$
reference frame relative to the direction of the kaon pair, (e) the proton
momentum in the $pp$ reference frame, and (f)
the $K^+K^-p$ invariant mass distribution.} \label{PPKK}
\end{figure}

Seven degrees of freedom are required to parametrize the unpolarized
$ppK^+K^-$ final state and these were chosen to be four angles, the
$K^+K^{-}$ and $K^+K^{-}p$ invariant masses, and the relative momentum of the
protons in the $pp$ system. Distributions in these seven variables were
generated inside the ANKE acceptance and compared with the experimental data
for non-$\phi$ data in Fig.~\ref{PPKK}. It is evident that the description of
the non-resonant $pp\to ppK^+K^-$ reaction at $T_{p}=2.83$~GeV is very
satisfactory and the same is true for the $K^+K^-$ invariant mass
distribution of Fig.~\ref{IM} when the $\phi$ contribution is added
incoherently.

Turning now to $\phi$ production in proton-proton collisions, the only
amplitude that survives at threshold corresponds to the
$^{3\!}P_1\,{\to}\,^{1\!}S_0s$ transition. We here denote the final state by
$^{2S+1}L_{J}\ell$, where $S$, $L$, and $J$ represent the total spin, orbital
angular momentum and total angular momentum of the $pp$ system, respectively,
and $\ell$ the orbital angular momentum of the $\phi$ relative to the $pp$
system. Our previous analysis indicates that the differential cross section
at an excess energy $\varepsilon_{\phi}= 18.5$~MeV is dominantly $S$-wave,
with a clear effect coming from the $pp$ final state
interaction~\cite{Har06}.

In contrast, significant contributions from higher partial waves were
suggested by the DISTO data at $\varepsilon_{\phi}=83$~MeV~\cite{Bal01},
where the differential cross sections as functions of the proton momentum in
the $pp$ reference frame and the momentum of $\phi$ meson in the c.m.\ system
were interpreted as reflecting the importance of $Ps$ and $Sp$ final waves,
respectively.  The anisotropy in the helicity distribution shows the
necessity also for a $Pp$ contribution.

There are several possible transitions that could lead to a $pp\phi$ final
state and we keep only typical ones in our model description, where the
spin-averaged squared transition matrix element is written as:
 \begin{eqnarray}
\nonumber\overline{|M|^2}&=&A_{Ss}\:(\hat{k}\times\hat{K})^2+
A_{Ps}\:\vec{p}^{\:2}+A_{Pp}(\vec{q}\cdot\vec{p}\,)^2\\
&&+A_{Sp}\left[3(\vec{q}\cdot\hat{K})^2-\vec{q}^{\:2}\right].
\label{model}
\end{eqnarray}
The momenta of the proton beam and $\phi$ meson in the overall c.m.\ system are
denoted by $\vec{K}$ and $\vec{q}$, respectively, $\vec{k}$ represents the
momentum of decay kaons in the $\phi$ reference frame, and $\vec{p}$ is the
relative momentum in the final $pp$ system.

\begin{figure}[h!]
\centering
\includegraphics[width=1.05\columnwidth,clip]{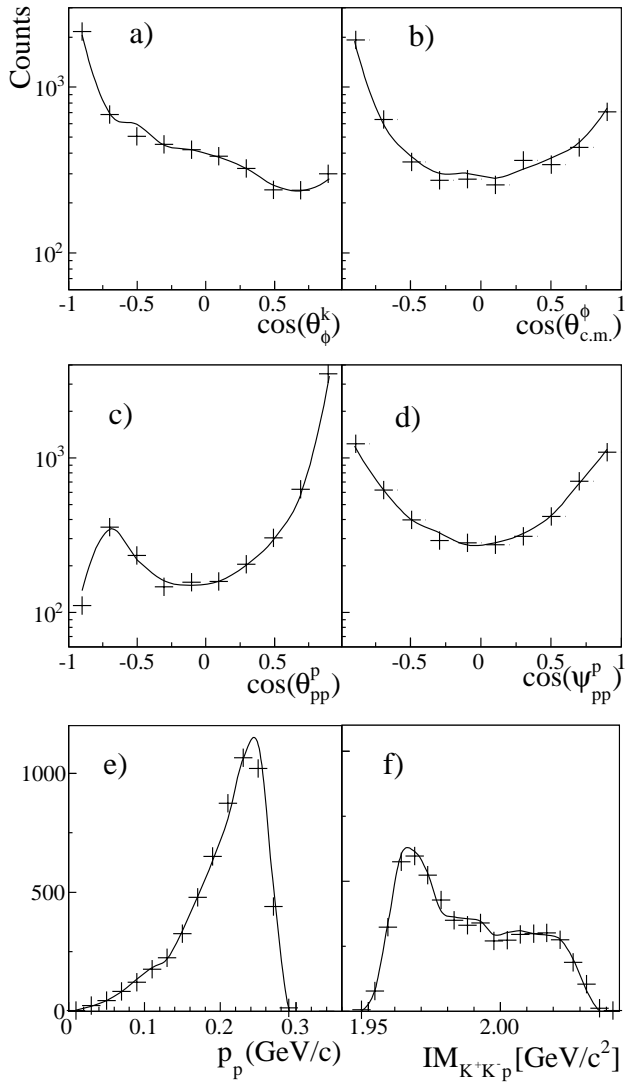}
\caption{The differential distributions of yields in the $\phi$ region
($1.01~\textrm{GeV}/c^2< \textrm{IM}_{K^+K^-}< 1.03~\textrm{GeV}/c^2$) for
the $pp\to ppK^+K^-$ reaction at $T_p=2.83$~GeV, where the points are
experimental and the curves represent simulations. The notations for the six
panels are the same as those in Fig.~\ref{PPKK}.} \label{PPPHI}
\end{figure}

Apart from the explicit momentum factors, we assume that the coefficients
$A_{L\ell}$ in Eq.~\eqref{model} are constant except that, at low invariant
masses, the final $pp$ system in the $^{1\!}S_{0}$ state is subject to a very
strong final state interaction. The $A_{Ss}$ and $A_{Sp}$ contributions in
Eq.~\eqref{model} were therefore multiplied by an enhancement factor which
was calculated using the Jost function,
\begin{equation}
\label{fsi} F_{pp}(q_{pp})=|J(q_{pp})|^{-2}=\frac{q_{pp}^2+\beta^2}{q_{pp}^2+\alpha^2}\,,
\end{equation}
where we take $\alpha=0.1$~fm$^{-1}$ and $\beta=0.5$~fm$^{-1}$~\cite{Mae08}. The
Coulomb interaction was neglected and, crucially, no attempt was made to include a
final state interaction in the $\phi p$ system.

A Monte Carlo phase-space simulation was written, based on
\textit{GEANT4}~\cite{GEANT4}, which took into account the detector
efficiency, resolution and kaon decay probability. Contributions from the
phenomenological parametrizations for $\phi$ and non-$\phi$ production were
then included as weights. The $\phi$ meson was taken to have a Breit-Wigner
form with a width of $\Gamma=4.26$~MeV/$c^2$~\cite{PDG10}, convoluted with a
resolution width of $\sigma \simeq 1$~MeV/$c^2$. The values of the
coefficients $A_{L\ell}$ in Eq.~\eqref{model} were determined by minimizing
$\chi^2$ in the difference between the simulated and experimental spectra,
and the results are shown in Table~\ref{fit}. The resulting descriptions of
the experimental data in Figs.~\ref{IM}, \ref{PPKK}, and \ref{PPPHI} are very
good and certainly adequate for carrying out the acceptance corrections.

\begin{table}[hbt]
\caption{\label{fit}%
Values of the model parameters of Eq.~\eqref{model} deduced by comparing the
simulations with data in the $\phi$ region. The momenta are measured in
GeV/$c$. All the parameters are normalized to $A_{Ss}=1$ and the
corresponding uncertainty of $\pm 0.25$ is not included.  }
\begin{center}
\begin{tabular}{|c|c|}
\hline
Parameter& Fit value\\ \hline
$A_{Ss}$ & \phantom{1}1.0 \\
$A_{Sp}$ & $\phantom{1}9.9\pm 1.8$\\
$A_{Ps}$ & $143\pm 4\phantom{1}$ \\
$A_{Pp}$ & $293\pm 21$ \\
\hline
\end{tabular}
\end{center}
\end{table}

The fits of the parametrizations to the experimental data allow the
extraction of the differential and total cross sections for both $\phi$ and
non-$\phi$ kaon pair production. The luminosity needed for this analysis was
determined on the basis of the $pp$ elastic scattering data that were taken
in parallel, using the forward detector~\cite{Mae08}. The associated
systematic uncertainty is estimated to be 7\%. Systematic uncertainties also
arise from the background subtraction, tracking efficiency, and the
model-dependent acceptance corrections. The latter were estimated from the
differences between the distributions corrected by the parametrization and
those corrected by phase space. As the observed distributions deviate
significantly from phase space, such estimates provide upper limits on these
uncertainties.

%
%

\section{Differential cross sections for $\boldsymbol{\phi}$ production}
\label{phiresults}

\begin{figure}[hbt]
\begin{center}
\includegraphics[width=1\columnwidth,clip]{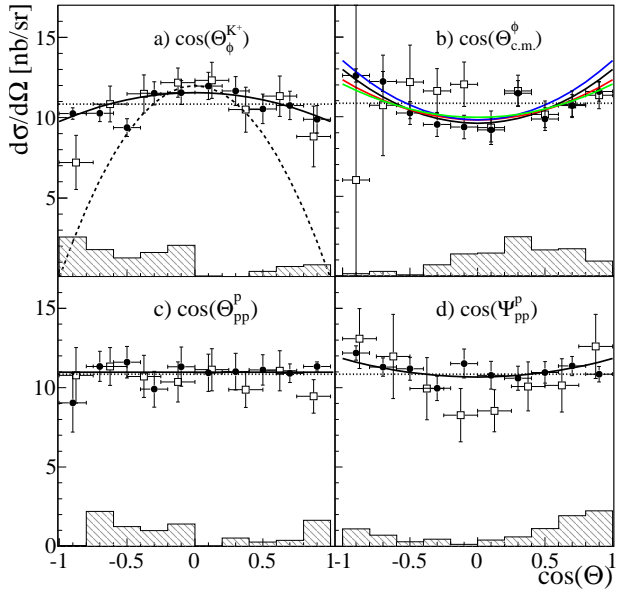}
\caption{(Color online) Angular distributions of the $pp\to pp\phi$ reaction
obtained in this experiment (solid circles), where the systematic
uncertainties are shown by the hatched histograms, compared with the scaled
DISTO data (open squares)~\cite{Bal01}. The dotted curves represent isotropic
distributions whereas the solid ones show fits to the ANKE results. (a) The
distribution with respect to the cosine of the $K^+$ polar angle in the
$\phi$ rest frame (Decay angle). The dashed curve demonstrates a
$\sin^2\theta^{K}_{\phi}$ behavior. (b) The distribution in the $\phi$ polar
angle in the overall c.m.\ system. The blue, red and green curves are typical
theoretical predictions from Refs.~\cite{Kap05},~\cite{Tsu03} and
\cite{Xie08}, respectively. (c) The distribution in the proton polar angle in
the $pp$ reference frame relative to the incident proton direction (Jackson
angle). (d) The distribution of the proton polar angle in the $pp$ reference
frame relative to the $\phi$ direction (Helicity angle). } \label{Target}
\end{center}
\end{figure}

The angular distributions for the $pp\to pp\phi$ reaction measured in this
experiment and that of DISTO~\cite{Bal01} are shown in Fig.~\ref{Target}.
These distributions must be symmetric about $\cos\theta=0$ and the data can
be parametrized in the form:
\begin{eqnarray}
\label{fiteq}
\frac{d\sigma}{d\Omega}= a\left[1 + b\,P_2(\cos\theta)\right].
\end{eqnarray}
The numerical values of the coefficients obtained from fitting the data are
reported in Table~\ref{fitresult}.

\begin{table}[h!]
\caption{\label{fitresult}  Values of the coefficients of Eq.~\eqref{fiteq}
for the $K^+$ decay angle with respect to the beam direction, the c.m.\
production angle, and the helicity angle, deduced by fitting the data of ANKE
and DISTO~\cite{Bal01}. The DISTO data have been scaled by 0.7 in order to
allow a direct comparison of the two sets of results.}
\begin{center}
\begin{tabular}{| l | c | c |  c | c | } \hline
 \multicolumn{1}{|c| }{} & \multicolumn{2}{|c| }{ ANKE  } & \multicolumn{2}{ |c| }{DISTO (scaled by 0.7)} \\ \cline{2 - 5}
  & $a$~[nb/sr] & $b$  & $a$~[nb/sr]  & $b$   \\ \hline
{cos~$\theta^K_{\phi}$} & {$10.96\pm0.23$} &{$-0.11\pm0.04$} &{$10.78\pm0.50$} &{$-0.27\pm0.08$} \\ 
{cos~$\theta^{\phi}_{c.m.}$} & {$10.71\pm0.21$} &{$\phantom{-}0.21\pm0.04$} &{$10.81\pm0.45$} &{$\phantom{-}0.07\pm0.07$} \\ 
{cos~$\Psi_{pp}^{p}$} & {$11.06\pm0.22$} &{$\phantom{-}0.07\pm0.04$} &{$10.64\pm0.64$} &{$\phantom{-}0.30\pm0.14$} \\ \hline
\end{tabular}
\end{center}
\end{table}

In the near-threshold region where the $Ss$ final state dominates the $\phi$
meson spin must lie along the beam direction. The polar angular distribution
of the decay kaons in the $\phi$ meson rest frame should then display a
$\sin^2\theta^{K}_{\phi}$ distribution, where $\theta^{K}_{\phi}$ is the
angle of a daughter kaon from the $\phi$ decay in the $\phi$ rest frame. The
data at $\varepsilon_{\phi}=18.5$~MeV~\cite{Har06} are  consistent with such
a dependence and deviations from this behavior are a sign of higher partial
waves.

Quite generally, the differential cross section is of the form:
\begin{equation} \label{decay} \frac{d\sigma}{d\Omega} \propto
\left[(1-\rho_{00})\sin^2\theta^{K}_{\phi}+2\rho_{00}\cos^2\theta^{K}_{\phi}\right],
\end{equation}
where $\rho_{00}= (1+b_K)/3$ is a spin density matrix element. From the value
of $b_K$ given in Table~\ref{fitresult} it is seen that the ANKE results
correspond to $\rho_{00}=0.30\pm0.01$, which is close to the unpolarized
value of $\frac{1}{3}$. This is to be compared with the value of
$\rho_{00}=0.23\pm0.04$ reported by DISTO at the marginally higher
$\varepsilon_{\phi}=83$~MeV~\cite{Bal01} where, in both cases, only
statistical errors are quoted. These are model-independent proofs that higher
partial waves are important at even relatively modest excess energies. A
similar conclusion is reached in a study of the available $pn\to d\phi$
data~\cite{Mae06}.

The angular distribution of the $\phi$ meson in the overall c.m.\ frame shown
in Fig.~\ref{Target}b is symmetric within experimental uncertainties. The
ANKE data show a stronger anisotropy than those of DISTO, as indicated by the
larger $b$ parameter in Table~\ref{fitresult}, but the error bars of the
DISTO data are significant. Theoretical calculations~\cite{Tsu03,Kap05,Xie08}
can describe a non-isotropy in the experimental data reasonably well, as
shown in Fig.~\ref{Target}b. The angular distribution is expected to be
isotropic when the mesonic current is dominant, whereas the nucleonic current
leads to a $\cos^2\theta$ distribution. The angular distribution might
therefore provide some information on the $NN\phi$ coupling
constant~\cite{Nak99,Tsu03}.

It could be interesting to compare our or the DISTO results of
Table~\ref{fitresult} with the analogous measurement at COSY-TOF of the
$pp\to pp\,\omega$ reaction at an excess energy of 92~MeV~\cite{Abd10}.
Unfortunately, the error bars in the $\omega$ angular distribution,
$1.0+(0.23\pm0.26)P_2(\cos\theta)$, are too large to draw any useful
conclusions as to whether the shapes are similar or not.

The distribution in the proton polar angle measured in the $pp$ reference
frame relative to the beam direction is nearly isotropic, as shown in
Fig.~\ref{Target}c. This is consistent with the DISTO results. On the other
hand, the analogous observable relative to the $\phi$ direction shown in
Fig.~\ref{Target}d has some anisotropy. This feature, which was also seen in
the DISTO data~\cite{Bal01}, is evidence for a contribution from a $Pp$ final
wave.

\begin{figure}[h!]
\centering
\includegraphics[width=0.9\columnwidth,clip]{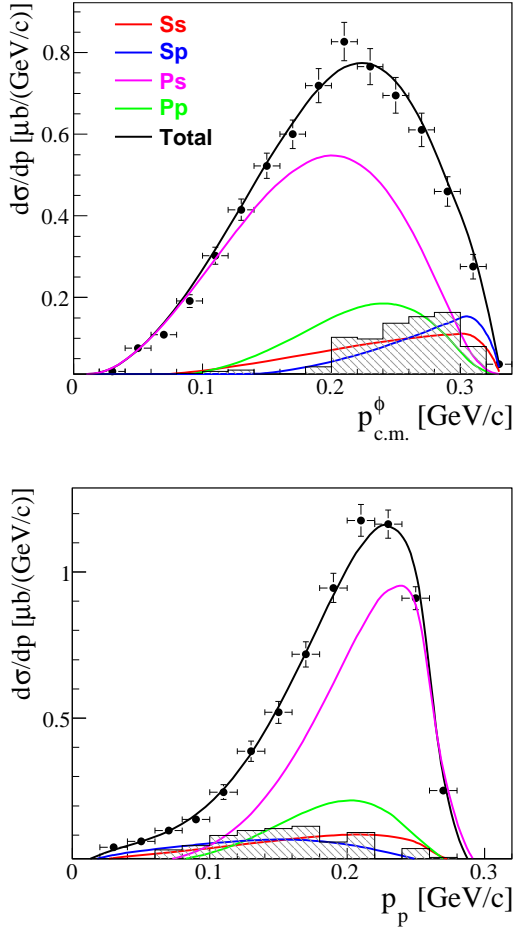}
\caption{(Color online) (Upper panel) Differential cross section for the
$pp\to pp\phi$ reaction as a function of the momentum of the $\phi$ meson in
the c.m.\ system. (Lower panel) Differential cross section for the $pp\to
pp\phi$ reaction as a function of the proton momentum in the $pp$ rest frame.
The systematic uncertainties are shown by the hatched histograms. The curves
represent the fitted contributions of different partial waves within the
parametrization of Eq.~\eqref{model}.} \label{mom}
\end{figure}

In neither the ANKE data at $\varepsilon_{\phi}=76$~MeV nor those of DISTO at
$\varepsilon_{\phi}=83$~MeV is there any sign of the FSI enhancement in the
proton-proton relative momentum spectrum. The lack of such an effect can be
understood by looking at the momentum distributions of the $\phi$ meson in
the c.m.\ system and relative momentum distribution of the final protons in
the $pp$ reference frame that are shown in Fig.~\ref{mom}. The contributions
of the different partial waves obtained by fitting Eq.~\eqref{model} to the
ANKE data are also indicated. From these it is seen that, within the given
parametrization, the $pp$ $P$-waves are completely dominant and this reduces
considerably the influence of the $^{1\!}S_{0}$ $pp$ FSI.

\begin{figure}[h!]
\centering
\includegraphics[width=1.0\columnwidth,clip]{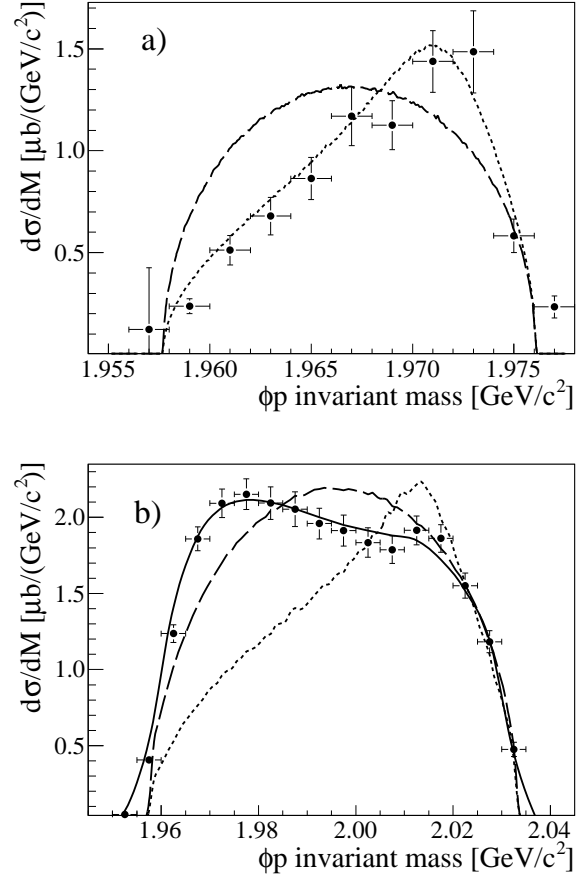}
\caption{ The acceptance-corrected differential cross section as a function
of the $\phi p$ invariant mass at excess energies (a) $\varepsilon_{\phi}=18.5$~MeV
and (b) $\varepsilon_{\phi}=76$~MeV. The dashed curves show phase-space predictions,
while the dotted cures include the $pp$ FSI. The solid curve represents the
description of Eq.~\eqref{model}, with parameters being taken from
Table.~\ref{fit}. } \label{IMKKp}
\end{figure}

The invariant mass distributions of the $\phi p$ system obtained in this
experiment and in the previous one at $\varepsilon_{\phi}=18.5$~MeV are
presented in Fig.~\ref{IMKKp}. For both energies the data differ
significantly from uniform phase-space predictions (dashed curve).
Calculations that include in addition the $pp$ final state interaction
(dotted curve) can describe the data at $\varepsilon_{\phi}=18.5$~MeV, but
fail at $\varepsilon_{\phi}=76$~MeV, where the higher partial waves of
Eq.~\eqref{model} are successful (solid curve).

%
%

\section{The total cross sections}
\label{total}

\begin{figure}[h!]
\centering
\includegraphics[width=1.0\columnwidth,clip]{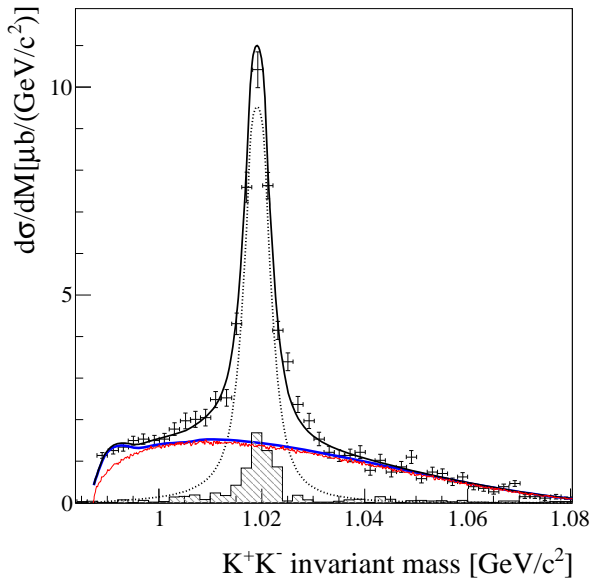}
\caption{(Color online) The acceptance-corrected $pp\to ppK^+K^-$
differential cross section as a function of the $K^+K^-$ invariant mass. The
error bars correspond only to the statistical uncertainties; systematic
uncertainties are shown by the hatched histograms. The blue curve shows the
non-$\phi$ contributions within the fitted parametrization, the red curve the
four-body phase-space simulation of $ppK^+K^-$, and the dotted histogram the
$\phi$ contributions. The solid line is the incoherent sum of the $\phi$ and
non-$\phi$ contributions. } \label{IMCO}
\end{figure}

The peaking at low IM$_{K^+K^-}$ in the raw $K^+K^-$ invariant mass
distribution of Fig.~\ref{IM} is mainly a consequence of the ANKE acceptance
and a smoother behavior in this region is seen in the acceptance-corrected
data in Fig.~\ref{IMCO}. The contributions are there shown separately for the
$\phi$ and non-$\phi$ contributions. Away from the low-mass region the latter
resembles quite closely that of a four-body $ppK^+K^-$ phase space, which is
also shown.

The fit to the acceptance-corrected invariant mass distribution of
Fig.~\ref{IMCO} has been used to determine separately the total cross
sections for $\phi$ and non-$\phi$ production measured in the $pp\to
ppK^+K^-$ reaction at 2.83~GeV. These results, together with our previous
ones at this energy, are summarized in Table~\ref{table1}. The two data sets
are consistent within statistics, though the precision of the current one is
much higher. It should be noted that the total cross section for $\phi$
production has been corrected for the branching ratio
$\Gamma_{K^{+}K^{-}}/\Gamma_{\text{tot}}=0.491$~\cite{PDG10}.

\begin{table}[h!]
\caption{\label{table1}%
Total cross section for the $pp\to ppK^{+}K^{-}$ reaction at $T_p=2.83$~GeV
separated into $\phi$ and non-$\phi$ components. In the $\phi$ case the data
have been corrected for the $\phi\to K^+K^-$ branching ratio. The
uncertainties are, respectively, statistical and systematic. The results of
previous measurements~\cite{Har06,Mae08} are also given.}
\begin{center}
\begin{tabular}{|c|c|c|}
\hline
Channel&$\sigma$[nb] & $\sigma$[nb]~{\cite{Har06,Mae08}}\\ \hline
Non-$\phi$ production &$\phantom{1}91.0\pm 3.0\pm 11.4$ & $\phantom{1}98.0\pm \phantom{1}8.0\pm 15.0$\\
$\phi$ production&  $142.2\pm2.1\pm17.9$ & $133.0\pm12.0\pm27.0$\\
Total $K^+K^-$& $160.8\pm3.2\pm14.4$ & $163.3\pm10.0\pm20.0$\\
\hline
\end{tabular}
\end{center}
\end{table}

\begin{figure}[h!]	
\centering
\includegraphics[width=1.0\columnwidth,clip]{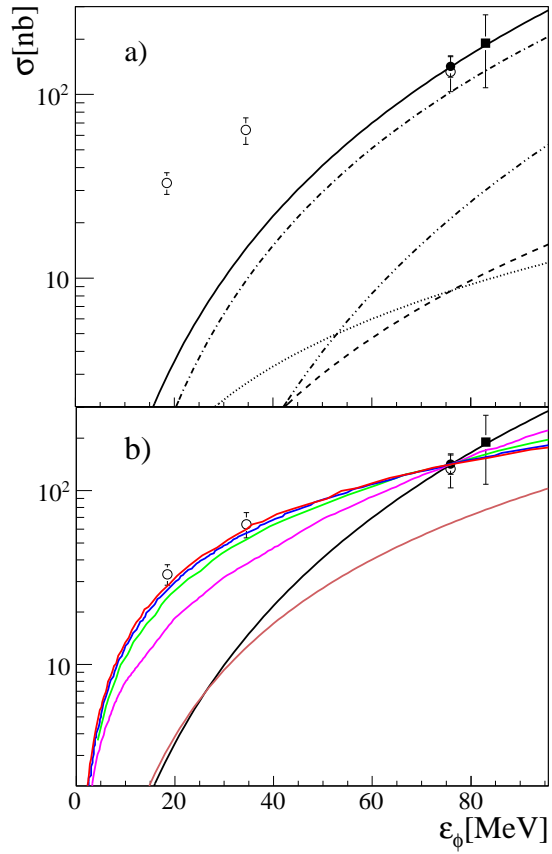}
\caption{(Color online) Total cross section for the $pp\to pp\phi$ reaction
as a function of excess energy $\varepsilon_{\phi}$. The present result (solid
circle) is shown together with experimental data taken from
DISTO~\cite{Bal01} (solid square) and previous ANKE measurements~\cite{Har06}
(open circles). a) The data are compared to the black solid curve derived using
Eq.~\eqref{model} with the parameters taken from Table.~\ref{fit}. The
individual contributions from the $Ss$ (dotted curve), $Sp$ (dashed curve),
$Ps$ (dashed-dotted curve), and $Pp$ (double dotted-dashed) are normalized to
their predicted values at 76~MeV.\\
b) The predictions of Tsushima and Nakayama~\cite{Tsu03} (magenta), scaled to
pass through the 76~MeV point, underestimate the low energy data. Also shown
are the predictions of Kaptari and K\"ampfer~\cite{Kap05} (green), which are
very similar to those of three-body phase-space with the inclusion of the
$pp$ FSI, the results within a resonance model Xie~\textit{et
al.}~\cite{Xie08} (blue), and a one-pion-exchange model of
Sibirtsev~\cite{Sib96} (brown), and this plus exotic baryons~\cite{Sib06}
(red).} \label{txs}
\end{figure}

The total cross section for the $pp\to pp\phi$ reaction is plotted in
Fig.~\ref{txs}a alongside other existing near-threshold
data~\cite{Har06,Bal01} as a function of the excess energy
$\varepsilon_{\phi}$. The error bars shown are quadratic sums of the
systematic and statistical uncertainties. If the coefficients $A_{L\ell}$
were constant, apart from the explicit momentum factors in Eq.~\eqref{model},
then these could be used to predict the energy dependence of the total cross
section. The resulting black solid curve, which by construction passes
through the 76~MeV point, underestimates severely the low energy data. This
behaviour comes about because at 76~MeV the fit indicates that only a small
fraction of the total cross section corresponds to a $Ss$ final state and, as
seen in Fig.~\ref{txs}a, the contributions from the higher partial waves
decrease faster as threshold is approached. It therefore seems that there
must be a strong energy variation in some of the $A_{L\ell}$, which might be
driven by a $\phi p$ near-threshold enhancement.

The energy dependence of the total cross section is close to the predictions
from Kaptari and K\"{a}mpfer~\cite{Kap05}, which include mesonic and
nucleonic current contributions. The predictions are very similar to those of
three-body phase space modified by the effects of the $pp$ FSI. This curve
can fit most of the data in Fig.~\ref{txs}b because, unlike the $Ss$ curve,
it takes the full strength at 76~MeV. The model of Tsushima and
Nakayama~\cite{Tsu03} also includes both nucleonic and mesonic current
contributions but gives too steep an energy dependence. In neither model were
contributions from nucleon resonances considered which, if they existed,
would change the energy dependence of the $A_{L\ell}$. Also shown are the
predictions of the resonance model of Xie \textit{et al.}~\cite{Xie08}. For
ease of comparison, these have all been scaled to pass through the 76~MeV
point. On the other hand, the one-pion-exchange calculation~\cite{Sib96},
which fits the $\phi$ production results at high energy ($\varepsilon_{\phi}
> 1$~GeV), fails to describe any of the near-threshold data. The model was
subsequently extended through the inclusion of baryonic resonances with
masses close to the $\phi p$ threshold~\cite{Sib06}. This achieves a better
description at lower energies, as shown in Fig.~\ref{txs}b. It is clear from
this discussion that the behavior of the total cross sections is insufficient
by itself to distinguish between different theoretical models; such
calculations must be tested against various differential spectra of the types
presented here.

The current value of the $pp\to pp\phi$ total cross section at 76~MeV given
in Table~\ref{table1} is only a little higher than our previous
result~\cite{Mae08}. The conclusion drawn there, that the ratio of this to
the cross section for $\omega$ production is about a factor of six above the
OZI limit~\cite{OZI}, is therefore still valid.

Values of the non-$\phi$ contribution to the $pp\to ppK^+K^-$ total cross
section were reported in our earlier work~\cite{Mae08} and any change in the
76~MeV point is well within the total error bars. It was shown there that the
energy dependence of this cross section could only be understood fully if all
the final state interactions in the $pp$, $K^-p$, and $K^+K^-$ subsystems
were included in the estimates.

%
%

\section{Non-$\boldsymbol{\phi}$ invariant mass distributions}
\label{nonphi}

\begin{figure}[h!]
\centering
\includegraphics[width=1.0\columnwidth,clip]{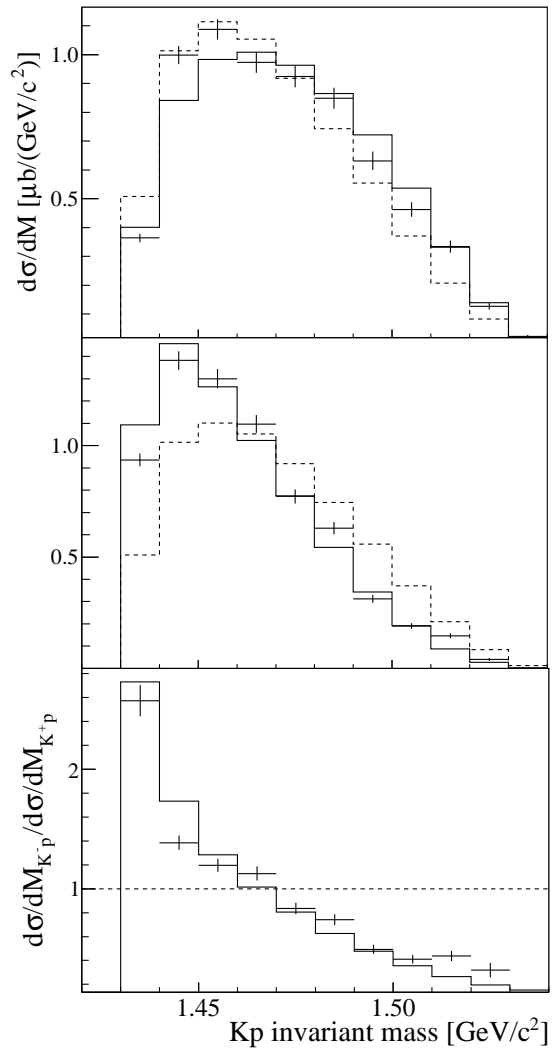}
\caption{Differential cross sections for the $pp\to ppK^+K^-$ reaction in the
non-$\phi$ region with respect to the invariant masses of $K^+p$ (upper
panel) and $K^-p$ (middle panel), and their ratio $R_{Kp}$ (lower panel). The
dashed histograms represent the four-body phase-space simulations, whereas
the solid ones represent the theoretical calculations taking into account
$pp$ and $K^-p$ final state interactions through Eq.~\eqref{assume}.}
\label{Exc_pK}
\end{figure}

\begin{figure}[h!]
\centering
\includegraphics[width=1.0\columnwidth,clip]{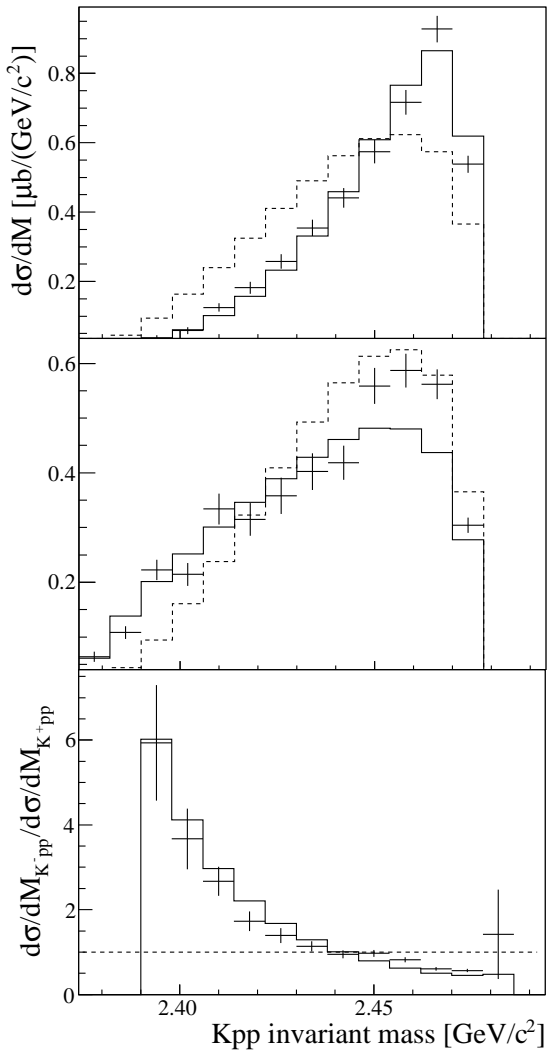}
\caption{Differential cross sections for the $pp\to ppK^+K^-$ reaction in the
non-$\phi$ region with respect to the invariant masses of $K^+pp$ (upper
panel) and $K^-pp$ (middle panel), and their ratio $R_{Kpp}$ (lower panel).
The dashed histograms represent the four-body phase-space simulations,
whereas the solid ones represent the theoretical calculations taking into
account $pp$ and $K^-p$ final state interactions through Eq.~\eqref{assume}.}
\label{Exc_ppK}
\end{figure}

The strong $K^-p$ interaction can distort hugely both the $K^-p$ and $K^-pp$
invariant mass distributions, and this is taken into account through
Eq.~\eqref{assume}. The effects are most apparent if one forms the ratios of
the differential cross sections in terms of the invariant masses:
\begin{equation}
\label{IMratio}
R_{Kp} = \frac{d\sigma/dM_{K^-p}}{d\sigma/dM_{K^+p}},\ \
R_{Kpp} = \frac{d\sigma/dM_{K^-pp}}{d\sigma/dM_{K^+pp}}\cdot
\end{equation}
The corresponding experimental data and simulations are to be found in
Figs.~\ref{Exc_pK} and \ref{Exc_ppK}. If the $K^+p$ and $K^-p$ final state
interactions were identical, then the ratios $R_{Kp}$ and $R_{Kpp}$ would be
constant and equal to unity. However, both $R_{Kp}$ and $R_{Kpp}$ display
very large preferences for lower invariant masses, which probably reflect an
attraction between the $K^-$ and one or both of the protons. Similar effects
have been observed at lower excess energies~\cite{Mae08,Win06,Sil09}.

The general features of these results are well reproduced by the simple
factorized ansatz of Eq.~\eqref{assume}. It is nevertheless surprising that
the distortions produced by the constant effective scattering length, $a=
(0+1.5i)$~fm, used at $\varepsilon_{KK} = 51$~MeV~\cite{Mae08} still describe
the data so well at an excess energy with respect to the $K^+K^-$ threshold
as high as 108~MeV, though some deviations are apparent for $Kp$ invariant
masses above about 1.5~GeV/$c^2$.

The distortions of Figs.~\ref{Exc_pK} and \ref{Exc_ppK} clearly indicate that
the direct production of the scalar resonance $a_0$ or $f_0$ cannot be the
dominant driving mechanism in the $pp\to ppK^+K^-$ reaction. On the other
hand, the strength of the $K^-p$ interaction suggests that kaon pair
production might be related to that of the $\Lambda(1405)$ through $pp\to
pK^+(\Lambda(1405)\to K^-p)$~\cite{Wil09}. This idea was put on a
quantitative footing by assuming that the $\Lambda(1405)$ was formed through
the decay $N^{\star}\to K^+\Lambda(1405)$~\cite{Xie10}.

The simple ansatz of Eq.~\eqref{assume} underestimates the cross section for
low $K^+K^-$ masses, \textit{i.e.}, in the interval between the $K^+K^-$ and
$K^{\circ}\bar{K}^{\circ}$ thresholds at 987.4 and 995.3~MeV/$c^2$,
respectively. Similar effects were observed in $pp\to ppK^+K^-$ by
DISTO~\cite{Bal01} and by ANKE in $pn\to dK^+K^-$~\cite{Mae09}. Although
these enhancements must be due to $K\bar{K}$ final state interactions,
including $K^+K^-\rightleftharpoons K^0\bar{K}^0$ charge exchange scattering,
they could be connected with some small production of the $a_0/f_0$ scalar
resonances. However, in reality, the data are only sensitive to the
$K\bar{K}$ scattering lengths.

\begin{figure}[h!]
\centering
\includegraphics[width=\columnwidth,clip]{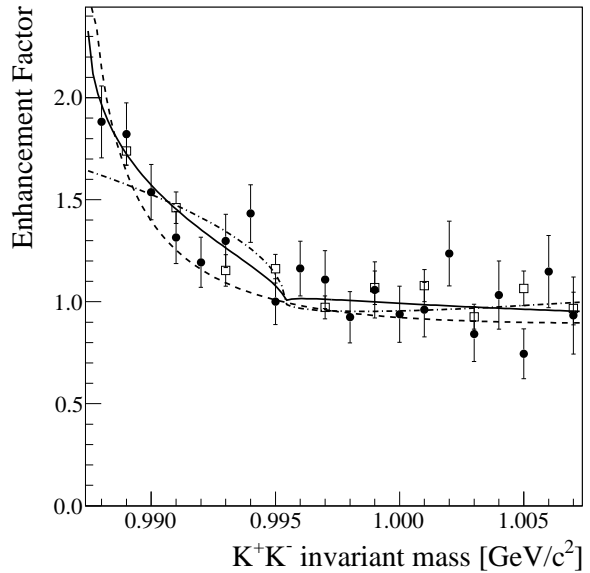}
\caption{Ratio of the measured $K^+K^-$ invariant mass in the $pp\to
ppK^+K^-$ reaction to estimates based on Eq.~\eqref{assume}. In addition to the
current data (solid circles), the weighted averages of previous measurements
(open squares)~\cite{Mae08} are also presented. The solid curve represents
the best fit in the model of Ref.~\cite{Dzy08}, which includes
charge-exchange and elastic $K^+K^-$ FSI. The best fits neglecting charge
exchange and including only this effect are shown by the dashed and the
dot-dashed curve, respectively.} \label{Enhance}
\end{figure}

A combined analysis of ANKE data at three energies~\cite{Dzy08} suggests
that, independent of the exact values of the scattering lengths, the
$K\bar{K}$ enhancement is mainly in the isospin-zero channel. The model for
the enhancement factor fitted there has been introduced into the simulation
to describe better the data shown in Fig.~\ref{IMCO} for invariant masses
IM$_{KK}< 995$~MeV/$c^2$. Its effects can be seen more clearly in the plot of
the ratio of the $K^+K^-$ invariant-mass data to the simulation based on
Eq.~\eqref{assume}, where no $K\bar{K}$ FSI was included. This, together with
the results of previous measurements~\cite{Mae08}, are shown in
Fig.~\ref{Enhance}. The two data sets are in agreement and are consistent
with the existence of some coupled-channel effect at the $K^0\bar{K}^0$
threshold but much better data would be required to prove this unambiguously.

%
%

\section{Discussion and Conclusions}
\label{conclusions}

New measurements of the differential and total cross sections for the
production of kaon pairs in proton-proton collisions have been presented at a
beam energy of 2.83~GeV. The reaction was identified through a triple
coincidence of a $K^+K^-$ pair and a forward-going proton detected in the
COSY-ANKE magnetic spectrometer, with an additional cut being placed on the
missing-mass spectrum.

By careful modeling, it was possible to describe all the experimental spectra
in regions of the $K^+K^-$ invariant mass where the $\phi$ meson sits, as
well as at smaller and larger masses. This allowed acceptance corrections to
be made in order to extract cross sections where the $\phi$ and non-$\phi$
contributions were reliably separated.

The main feature of the non-$\phi$ data is the very strong distortion of both
the $K^-p$ and $K^-pp$ spectra by the $K^-p$ final state interaction. This
may be a reflection of the excitation of the $\Lambda(1405)$ in the
production process and, in the Xie and Wilkin approach~\cite{Xie10}, the
production of non-$\phi$ kaon pairs proceeds mainly through the excitation of
$K^+$-hyperon pairs. It is remarkable to note that these distortions are
described quantitatively by the factorized approximation of
Eq.~\eqref{assume} with the same constant scattering length that was used for
the lower energy data~\cite{Mae08}. On the other hand, because these data
correspond to events where the $K^+K^-$ emerge in the final state, they
cannot contribute directly to the ongoing debate regarding the possibility of
deeply bound $K^-pp$ states~\cite{TYam10}, except to emphasize that the
$K^-pp$ interaction is still strong even above threshold.

There is evidence for some $K^+K^-$ final state interaction that changes in
nature at the $K^0\bar{K}^0$ threshold but the contribution of this region to
the integrated $pp\to ppK^+K^-$ cross section is very small and it is hard to
find any indication of the excitation of the $a_0/f_0$ scalar resonances in
the reaction. As already pointed out in our earlier work~\cite{Mae08}, the
energy dependence of the total cross section near threshold can be understood
simply in terms of the effects of the $pp$, $K^-p$, and $K^+K^-$ FSI. To
establish a better understanding of possible structure at the $K^0\bar{K}^0$
threshold, high statistics are required in this region and this might be
achieved in the data collected below $\phi$ production
threshold~\cite{Har08}.

Having a good description of the background, it was possible to derive
detailed invariant mass and angular distributions for the $pp\to pp\phi$
reaction. Although the DISTO collaboration~\cite{Bal01} showed the
significance of higher partial waves at the marginally higher excitation
energy of $\varepsilon_{\phi} = 83$~MeV, they did this mainly on the basis of
relative momentum spectra. Their conclusion is confirmed unambiguously by the
angular distributions presented here. For example, at
$\varepsilon_{\phi}=18.5$~MeV the $\phi$-meson is completely aligned, as it
has to be for a $Ss$ final state~\cite{Har06}. In contrast, in the present
data the emerging $\phi$ is almost unpolarized and this clearly signals the
presence of higher partial waves. This is consistent with the evidence from
the momentum distributions, which also show the dominance of $P$ waves in the
final $pp$ system. This explains why the $^{1\!}S_{0}$ $pp$ FSI, which is so
important at $\varepsilon_{\phi}=18.5$~MeV~\cite{Har06}, is not observed at
76~MeV. Furthermore, in contrast to the DISTO result~\cite{Bal01}, clear
anisotropy was observed in the $\phi$ c.m.\ angular distribution and this can
be ascribed to the contribution from $p$ wave. This angular distribution
might provide information on nucleonic current contributions and the $NN\phi$
coupling constant~\cite{Nak99,Tsu03}.

Even if one considers only a few partial waves, there are simply too many
parameters to perform useful fits and only typical $Ss$, $Sp$, $Ps$, and $Pp$
contributions were considered in Eq.~\eqref{model}. The fitted data show that
the contribution of the final $Ss$ wave to the cross section represents only
a small amount of the total at $\varepsilon_{\phi} = 76$~MeV. As a
consequence, the extracted parameters predict a total cross section that
grossly underestimates the measurements at lower energies.

The simplest way out of the total cross section dilemma would be to assume
that a $\phi p$ threshold enhancement leads to a significant energy
dependence of some of the $A_{L\ell}$ coefficients. In this context it is
interesting to note that the large contribution of the $Pp$ wave to the
$pp\to pp\eta$ cross section at an excess energy of 72~MeV was ascribed to a
strong $\eta p$ FSI driven by the $N^*(1535)$ isobar~\cite{PET2010}. Against
the $\phi p$ enhancement hypothesis is the fact that the large excess of
events in the $\phi p$ invariant mass distribution shown in Fig.~\ref{IMKKp}b
at low masses can be explained in the partial wave fitting of
Eq.~\eqref{model}, without including any $\phi p$ enhancement. We have not,
however, shown that the fitting of the data is unambiguous and there could be
other truncated partial wave forms that might be equally successful.
Furthermore, from the start we have not included any final state interaction
between $\phi$ and protons in the parametrization. There could therefore be a
possible trade-off between some of the partial wave parameters and an FSI in
the $\phi p$ system. Nevertheless, the phenomenological parametrization is
sufficient for acceptance correction and it describes well most of the
differential distributions.

In the parametrization of Eq.~\eqref{model}, the coefficients $A_{L\ell}$
were taken to be constant and no resonance effects were included. Recent
theoretical studies have suggested that bound states or resonances might be
formed in the near-threshold $\phi p$ system~\cite{Gao01,Yam10} and, if so,
they would certainly influence the behaviour of some of the $A_{L\ell}$. In
this context, it is interesting to note that a bump was observed in the
near-threshold $\phi$ meson photoproduction from hydrogen by
LEPS~\cite{Mib05} and in the preliminary results of CLAS~\cite{Dey11}.
Furthermore, it seems that $s$-wave production of the $\phi$ in the
$pd\to\,^3$He$\,X$ reaction is anomalously large compared to the $\omega$ and
$\eta'$ mesons~\cite{Fal95}. Such effects might even be part of the
explanation for the violation of the OZI rule~\cite{OZI} in the ratio of
$\phi$ to $\omega$ production.  Alternatively, it is possible that other
strangeness production channels could influence the energy dependence of the
$pp\to ppK^+K^-$ reaction~\cite{Dey10,Per10,Koh10}.

Although some theoretical models have been able to describe \textit{a
posteori} the published total cross sections for $\phi$ production,
calculations of differential distributions with which to compare our
experimental data are rather limited. It is only when a model is tested
against a range of differential distributions, as presented here, that some
credence can be given to the model. Total cross sections are insufficient and
more theoretical work is therefore required.

%
\begin{acknowledgments}
We would like to thank the COSY machine crew for their continued assistance
as well as that of other members of the ANKE Collaboration. Discussions with
J.~Haidenbauer, C.~Hanhart, K.~Nakayama and A. Sibirtsev were very helpful. This work has
been supported in part by the US Department of Energy under Contract
No.~DE-FG02-03ER41231, the BMBF, DFG, Russian Academy of Sciences, and COSY FFE.
\end{acknowledgments}
%
%

\end{document}